\documentclass[aps, prb, amsmath, twocolumn, 10pt, superscriptaddress]{revtex4-1}

\usepackage{hyperref}
\usepackage{graphicx}

\usepackage{braket}
\usepackage{bbold}
\usepackage{mathtools}

\begin{document}

\title{Vanishing Zeeman energy in a two-dimensional hole gas}

\author{P. Del Vecchio}
\thanks{These authors contributed equally to this work}
\affiliation{Department of Engineering Physics, \'Ecole Polytechnique de Montr\'eal, Montr\'eal, C.P. 6079, Succ. Centre-Ville, Montr\'eal, Qu\'ebec, Canada H3C 3A7}
\author{M. Lodari}
\thanks{These authors contributed equally to this work}
\affiliation{QuTech and Kavli Institute of Nanoscience, TU Delft, P.O. Box 5046, 2600 GA Delft, The Netherlands}
\author{A. Sammak}
\affiliation{QuTech and Netherlands Organisation for Applied Scientific Research (TNO), \\Stieltjesweg 1, 2628 CK Delft, The Netherlands}
\author{G. Scappucci}
\email{g.scappucci@tudelft.nl}
\affiliation{QuTech and Kavli Institute of Nanoscience, TU Delft, P.O. Box 5046, 2600 GA Delft, The Netherlands}
\author{O. Moutanabbir}
\email{oussama.moutanabbir@polymtl.ca}
\affiliation{Department of Engineering Physics, \'Ecole Polytechnique de Montr\'eal, Montr\'eal, C.P. 6079, Succ. Centre-Ville, Montr\'eal, Qu\'ebec, Canada H3C 3A7}

\begin{abstract}
A clear signature of Zeeman split states crossing is observed in Landau fan diagram of strained germanium two-dimensional hole gas. The underlying mechanisms are discussed based on a perturbative model yielding a closed formula for the critical magnetic fields. These fields depend strongly on the energy difference between the top-most and the neighboring valence bands and are sensitive to the quantum well thickness, strain, and spin-orbit-interaction. The latter is a necessary feature for the crossing to occur. This framework enables a straightforward quantification of the hole-state parameters from simple measurements, thus paving the way for its use in design and modelling of hole-based quantum devices.
\end{abstract}

\maketitle

\section{Introduction}

The inherently large and tunable spin-orbit interaction (SOI) energies of holes and their reduced hyperfine coupling with nuclear spins are behind the surging interest in hole spin qubits with fast all-electrical control.\cite{Hendrickx2020,Maurand2016,Watzinger2018,Scappucci2020,Hu2012} Holes can also host superconducting pairing correlations, a key ingredient for the emergence of Majorana zero modes\cite{Kloeffel2011,Maier2014a,Mao2012,Maier2014,Lutchyn2018} for topological quantum computing. Because of its attractive properties,\cite{Hendrickx2020,Watzinger2016,Sammak2019,Moutanabbir2010,Miyamoto2010,Bulaev2007,Wang2019,Hendrickx2018,Mizokuchi2018,Hendrickx2019,Vigneau2019,Gao2020,Lawrie2020} strained Ge low-dimensional system has been proposed as an effective building block to develop these emerging quantum devices. Interestingly, the simplicity of this system makes it a textbook model to uncover and elucidate subtle hole spin-related phenomena leading, for instance, to the recent observation of pure cubic Rashba spin-orbit coupling.\cite{Moriya2014}

Measuring Zeeman splitting (ZS) of hole states under an external magnetic field has been central in probing hole spin properties, as it is directly related to the hole g-factor, which is itself strongly influenced by the underlying SOI, strain, symmetry, and confinement.\cite{Kotlyar2001,Winkler2003} In III-V semiconductors,\cite{Kotlyar2001,Traynor1997,Lawless1992,Warburton1993,Jovanov2012,Danneau2006,Kubisa2011,Grigoryev2016,Fischer2007,FariaJunior2019,Tedeschi2019,Bardyszewski2014,Broido1985,Ekenberg1985} hole spin splitting depends nonlinearly on the out-of-plane magnetic field strength $B$, causing Landau level crossings/anti-crossings\cite{Warburton1993,Moriya2013} and Zeeman crossings/anti-crossings.\cite{Sammak2019,Lodari2019,Winkler1996} The nonlinearity is usually modeled by a quadratic-in-field contribution to ZS,\cite{Kotlyar2001} which owes its existence to valence band mixing. Depending on the sign of the splitting, Zeeman energy can even vanish at some finite critical field, $B_c$. Theoretical studies attribute these nonlinearities to the mixing of heavy-hole (HH) and light-hole (LH) bands at finite energy.\cite{Traynor1997} Alongside with valence band mixing, Rashba and Dresselhaus spin-orbit coupling were also shown to influence the crossing field, due to the lattice inversion asymmetry and the confining potential.

Detailed mechanisms of ZS of hole states are yet to be unravelled and understood and furthermore, ZS treatments for zinc-blende or diamond crystals that explicitly consider strain and SOI strength remain conspicuously missing in literature. Note that in early calculations\cite{Winkler1996} of Landau levels in Ge/SiGe quantum well (QW) to interpret cyclotron resonance experiments in Ref.~\onlinecite{Engelhardt1994}, the crossing of spin split states within the first HH subband was present and the corresponding field position was found to be sensitive to the strength of spin-orbit coupling. In that work, the authors insisted on the importance of including explicitly the split-off hole band, which was required to achieve a good agreement with experiments. Crucially, studies that included both strain and SOI were diagonalizing numerically the full $k\cdot p$ matrix.\cite{Jovanov2012,Winkler1996} However, this mathematical rigor comes at the expense of identifying the physics governing the non-linearities in ZS. 

To overcome these limitations and elucidate the underlying mechanisms of ZS, herein we uncover the clear signature of ZS crossings in a Ge high-mobility two-dimensional hole gas (2DHG). We also derive a theoretical framework describing the crossing of Zeeman split states that includes explicitly the SOI strength and strain. A closed formula for the crossing fields is obtained and validated by experiment. In addition to establishing the key parameters in Zeeman crossings, this analysis also provides a toolkit for a direct quantification from simple magnetotransport measurements of important physical quantities including HH out-of-plane g-factor, HH-LH splitting, and cubic Rashba spin-orbit coefficient.

\section{Experimental details}

The investigated 2DHG consists of a Ge/SiGe heterostructure including a strain-relaxed Si$_{0.2}$Ge$_{0.8}$ buffer setting the overall lattice parameter, a compressively-strained Ge QW, and a Si$_{0.2}$Ge$_{0.8}$ barrier separating the QW from a sacrificial Si cap layer. The growth was carried out in an Epsilon 2000 (ASMI) reduced pressure chemical vapor deposition reactor on a $100\,\text{mm}$ n-type Si(001) substrate. The growth sequence starts with the deposition of a Si$_{0.2}$Ge$_{0.8}$ virtual substrate. This virtual substrate is obtained by growing a $1.6\,\mu\text{m}$ strain-relaxed Ge buffer layer, a $1.6\,\mu\text{m}$ reverse-graded Si$_{1-x}$Ge$_x$ layer with final Ge composition $x = 0.8$, and a $500\,\text{nm}$ strain-relaxed Si$_{0.2}$Ge$_{0.8}$ buffer layer. A $16\,\text{nm}$ compressively-strained Ge quantum well is then grown on top of the Si$_{0.2}$Ge$_{0.8}$ virtual substrate, followed by a strain-relaxed $17\,\text{nm}$-thick Si$_{0.2}$Ge$_{0.8}$ barrier. An in-plane compressive strain $\epsilon_\parallel = -0.63\%$ is found in the QW via X-ray diffraction measurements.\cite{Sammak2019} A thin ($<2\,\text{nm}$) sacrificial Si cap completes the heterostructure. This cap is readily oxidized upon exposure to the cleanroom environment after unloading the Ge/SiGe heterostructure from the growth reactor.

\begin{figure}[t!]
    \centering
    \includegraphics[scale=0.8]{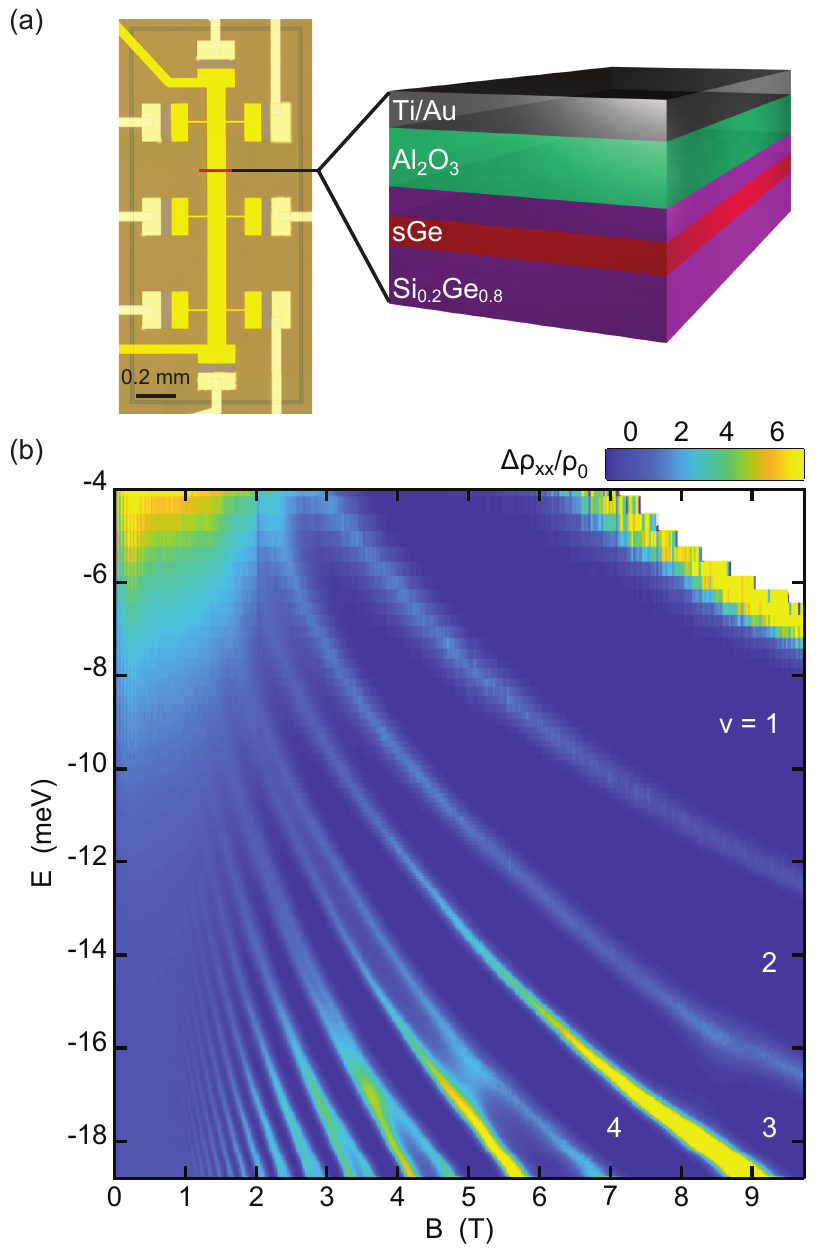}
    \caption{(a) Optical micrograph of a Hall-bar shaped Ge/SiGe heterostructure field effect transistor and cross section of the gate stack and active regions of the strained Ge/SiGe heterostructure below the red cut. The strained Ge (sGe) quantum well is $16\,\text{nm}$ thick and the Si$_{0.2}$Ge$_{0.8}$ barrier on top is $17\,\text{nm}$ thick. (b) Landau level fan diagram reporting the magnetoresistance $\Delta\rho_{xx}/\rho_0 = (\rho_{xx}-\rho_0)/\rho_0$ as a function of out-of-plane magnetic field $B$ and energy $E$. Labels of filling factors $\nu = 1$ -- $4$ are shown.}
    \label{fig:exp-fan-chart}
\end{figure}

Hall-bar field effect transistors (H-FETs) are fabricated and operated with a negatively biased gate to accumulate a 2D hole gas into the QW and tune the carrier density. Fig.~1a shows an optical micrograph of the H-FET and a cross-section schematic of the active layers and the gate stack. A $170\,\text{nm}$ deep trench mesa is dry-etched around the Hall-bar shaped H-FET in order to isolate the bonding pads from the device. The sample is dipped in HF to remove the native oxide prior to a $60\,\text{nm}$ Pt layer deposition via e-beam evaporation. Ohmic contacts are obtained by diffusion of Pt into the quantum well occurring during the atomic layer deposition of a $30\,\text{nm}$ Al$_2$O$_3$ dielectric layer at a temperature of $300\,^\circ\text{C}$. Finally, a $10/200\,\text{nm}$-thick Ti/Au gate layer is deposited. An optimized Si$_{0.2}$Ge$_{0.8}$ barrier thickness of $17\,\text{nm}$ was chosen, which is thin enough to allow for a large saturation carrier density\cite{Sammak2019} (up to $7.5\times 10^{11}\,\text{cm}^{-2}$), while providing sufficient separation to reduce scattering of carriers in the QW from remote impurities,\cite{Lodari2019} leading to large hole maximum mobility ($2.6\times 10^5\,\text{cm}^{-2}$). Large density range and high mobility are key ingredients to observe Landau level fan diagrams in magnetotransport with the clarity required to reveal subtle spin-related features.
 
In the magnetotransport studies, the longitudinal and transversal ($\rho_{xx}$ and $\rho_{xy}$) component of the 2DHG resistivity tensor were measured via a standard four-probe low-frequency lock-in technique. The measurements are recorded at a temperature of $T = 260\,\text{mK}$, measured at the cold finger of a $^3$He dilution refrigerator. A source-drain voltage bias $V_{sd} = 0.1\,\text{mV}$ is applied at a frequency of $7.7\,\text{Hz}$. The magnetoresistance characterization of the device is performed by sweeping the voltage gate $V_g$ and stepping $B$ with a resolution of $15\,\text{mV}$ and $25\,\text{mT}$, respectively. The energy $E$ is obtained using the relation $E = p\pi\hbar^2 / m^*$, where we obtain the carrier density $p$ by Hall effect measurements at low $B$ and we use the effective mass $m^*$ measured as a function of density in similar heterostructures.\cite{Lodari2019} The $\rho_{xx}$ vs. energy profiles in the upper panels of Figs.~3(a)-3(d) have been smoothed for clarity by using a Matlab routine based on Savitzky-Golay filtering method.

\section{Magnetotransport studies of strained Ge 2DHG}

The fan diagram in Fig.~1b shows the normalized magnetoresistance oscillation amplitude $\Delta\rho_{xx}/\rho_0 = (\rho_{xx} - \rho_0) / \rho_0$ as a function of energy and out-of-plane external magnetic field $B$ aligned along the growth direction $\mathbf{\hat{z}}$ and perpendicular to the 2DHG plane, where $\rho_0$ is the $\rho_{xx}$ value at $B = 0$. The Zeeman split energy gap, corresponding to odd integer filling factors $\nu$, deviates from its linear dependence on $B$, vanishes when the magnetic field reaches a critical value $B_c$, and then reopens at higher $B$ values. We clearly observe the associated crossing of Zeeman split states for odd integers $\nu = 3, 5, 7$, and $9$. Partial signatures of Zeeman crossings occurring at similar magnetic fields were observed in earlier studies,\cite{Sammak2019,Lodari2019} albeit the fan diagram measurements were limited in density range\cite{Sammak2019} or affected by thermal broadening.\cite{Lodari2019} These observations point to an underlying mechanism that is independent of the QW position with respect to the surface gate.

\section{Theoretical framework for hole dispersion in strained Ge 2DHG}

To identify the mechanisms behind the non-linearities in ZS and the parameters affecting the crossing field, we developed a perturbative model to describe the hole dispersion as a function of the out-of-plane magnetic field. The model assumes an abrupt and infinite band offset between the QW and its barriers and is based on a 6-band $k\cdot p$ Hamiltonian for HH, LH and split-off (SO) bands. The total Hamiltonian $H$ for the hole dispersion is written as\cite{Eissfeller2011}~: $H = H_k + H_\epsilon + H_{\text{SO}} + H_B + V$ where $H_k$ is a function of the wavevector operator $\mathbf{k} = (k_x,k_y,k_z)$, $H_\epsilon$ is the Bir-Pikus Hamiltonian and depends on the strain tensor components $\epsilon_{ij}$, $H_{\text{SO}}$ is the spin-orbit term proportional to the spin-orbit energy $\Delta$ and $H_B$ includes the interaction of the free electron spin with the magnetic field. $V$ is the infinite well potential for a square well of width $L$. We consider QWs grown along [001] direction and subjected to biaxial bi-isotropic strain. Thus, $\epsilon_{ij} = 0$ if $i\neq j$, $\epsilon_{xx} = \epsilon_{yy}\equiv \epsilon_\parallel$ and $\epsilon_{zz}=-D_{001}\epsilon_\parallel$, where $D_{001}$ is the Poisson ratio and $\epsilon_\parallel$ is the in-plane lattice strain.

We first rewrite the total Hamiltonian $H$ in two terms~: $H = H_0(\epsilon_\parallel; k_z) + H'(n, B; k_z)$, where the integer $n\geq 1$ labels the spin-split Landau pairs such that $\nu = 2n-1$ at crossings. The eigenstates of $H_0$ consist of pure HH subbands of energy $E_l^{\text{HH}}$ and two superpositions of LH and SO holes of energy $E_l^\eta$. Here, $\eta = \{+, -\}$ is a generic label to distinguish the two orthogonal LH-SO states and $l\geq 1$ is the subband index. The perturbation $H'$ introduces the magnetic field and is eliminated to second order by a Schrieffer-Wolff transformation, resulting in an effective Hamiltonian for the 2-fold HH subband. Remarkably, the resulting effective $2\times 2$ Hamiltonian for the HH subband does not couple spin-up ($+$) and spin-down ($-$) projections. The HH dispersion as a function of $B$ is thus simply the diagonal entries of the effective matrix. We have

\begin{align}
    \begin{split}\label{up}
      E_{+,l,n}^{(2)}(B) &= E_l^{\text{HH}} + 3n(n + 1)\left(\kappa - F_l\right)\frac{\mu_{\text{B}}B^2}{B_l^*}\\
      &- \left[\left(2n-1\right)\left(\gamma_1 + \gamma_2\right) + 3\kappa - 6nF_l\right]\mu_{\text{B}}B, \\
      \end{split}
\end{align}

\begin{align}
    \begin{split}\label{down}
      &E_{-,l,n}^{(2)}(B) = E_l^{\text{HH}} + 3(n - 2)(n - 1)\left(\kappa - F_l\right)\frac{\mu_{\text{B}}B^2}{B_l^*}\\
      &- \left[\left(2n-1\right)\left(\gamma_1 + \gamma_2\right) - 3\kappa - 6(n-1)F_l\right]\mu_{\text{B}}B, \\
    \end{split}
\end{align}

\noindent with

\begin{subequations}\label{B*}
\begin{align}
  B_l^* &= \frac{\kappa - F_l}{\mu_{\text{B}}\left(\gamma_2 + \gamma_3\right)^2}\left[\sum_{\eta = \pm}{\frac{\left(l_l^\eta + \sqrt{2}s_l^\eta\right)^2}{E_l^{\text{HH}} - E_l^\eta}}\right]^{-1}\label{B*_exact} \\
  &\approx \frac{\kappa - F_l}{\mu_{\text{B}}\left(\gamma_2 + \gamma_3\right)^2}\left(E_l^{\text{HH}} - E_l^{\text{LH}}\right)\label{B*_approx}
\end{align}
\end{subequations}

\noindent and

\begin{subequations}\label{F}
\begin{align}
  F_l &= \frac{32\alpha_0\gamma_3^2}{L^2}\sum_{\substack{j = 1 \\ j \neq l}}^\infty{\frac{\left[1-(-1)^{l+j}\right]l^2j^2}{\left(l^2 - j^2\right)^2}\sum_{\eta = \pm}{\frac{\left(l_j^\eta - s_j^\eta / \sqrt{2}\right)^2}{E_l^{\text{HH}} - E_j^\eta}}}\label{F_exact} \\
  &\approx \frac{32\alpha_0\gamma_3^2}{L^2}\sum_{\substack{j = 1 \\ j \neq l}}^\infty{\frac{1}{E_l^{\text{HH}} - E_j^{\text{LH}}}\frac{\left[1-(-1)^{l+j}\right]l^2j^2}{\left(l^2 - j^2\right)^2}}.\label{F_approx}
\end{align}
\end{subequations}

Here, $\mu_\text{B}$ is the Bohr magneton, $\gamma_i$ and $\kappa$ are the Luttinger parameters, $\alpha_0 = \hbar^2 / (2m_0)$ with $m_0$ the free electron mass and $l_l^\eta$ and $s_l^\eta$ are respectively the LH and SO contributions of the $l$th $\eta$ subband. The characteristic field $B_l^*$ controls the crossing positions and is filling factor-independent, while $F_l$ indicates the coupling strength between the HH subband and neighboring $\eta$ states. As we focus on the HH ground subband ($l = 1$), $l$ subscripts will be omitted for simplicity. The obtained Zeeman splitting energy $E_\text{Z}\equiv E_{-,n}^{(2)}(B) - E_{+,n}^{(2)}(B)$ of the $n$th spin-split Landau pair is~:

\begin{equation}\label{zeeman}
    E_\text{Z} = 6(\kappa - F)\mu_\text{B}B\left[1-(2n-1)\left(\frac{B}{B^*}\right)\right].
\end{equation}

\begin{figure*}[t!]
    \centering
    \includegraphics[scale=0.8]{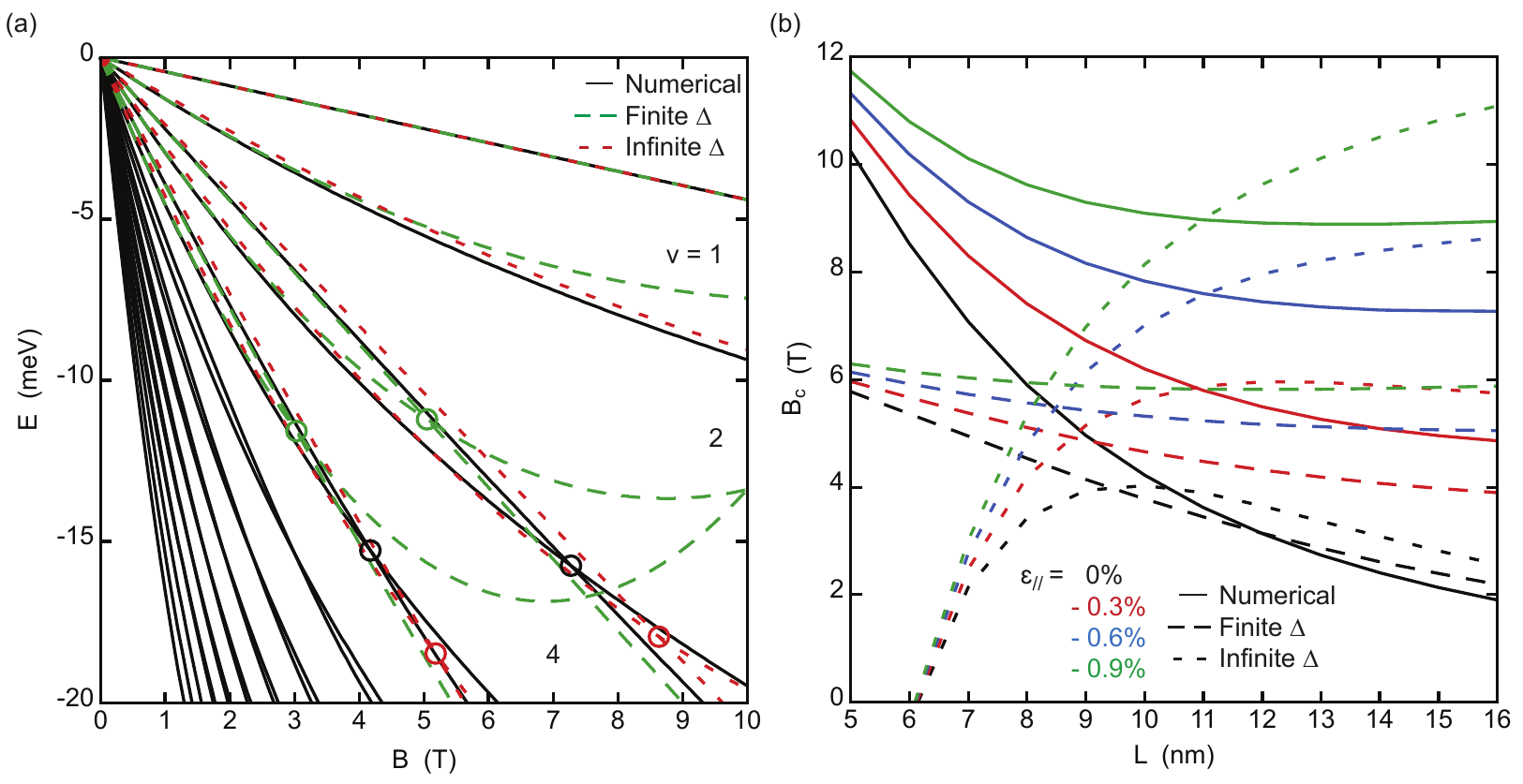}
    \caption{(a) Fan diagram of the ground HH subband in a $16\,\text{nm}$ Ge well subject to $0.6\%$ compressive strain. Solid curves are the dispersion obtained from the numerical solution of $H$, while the dashed curves are obtained from the second order dispersion assuming finite or infinite SOI respectively. Circles indicate the Zeeman crossings. Filling factors $\nu$ are also indicated. (b) $\nu = 3$ crossing field as a function of the well thickness at various strain values obtained from the numerical solution of $H$ (solid curves) and through Eq. \eqref{crossing} assuming finite or infinite SOI (dashed curves).}
    \label{fig:theory}
\end{figure*}

Solving for $E_\text{Z} = 0$ results in a second order approximation for the filling factor-dependent crossing field $B_c$~:

\begin{equation}\label{crossing}
  B_c^{(2)}(n) = \frac{B^*}{2n-1}.
\end{equation}

The energy difference that separates the HH subband edge from the energy at a crossing position can also be found from the second order equations. When $n\to\infty$ (or $\nu\to\infty$) this energy difference is independent of $n$~:

\begin{equation}\label{deltaE}
  \Delta E = \left[\gamma_1 + \gamma_2 - \frac{3}{4}\left(\kappa + 3F\right)\right]\mu_{\text{B}}B^*.
\end{equation}

Equation \eqref{zeeman} also yields the HH weak-field g-factor~:

\begin{equation}\label{g-factor}
  g^* = 6(\kappa - F).
\end{equation}

The approximations \eqref{B*_approx} and \eqref{F_approx} hold only when SOI is large enough so that the SO band can be neglected from the $k\cdot p$ framework. An explicit criterion for this is (Appendix \ref{app:H0})~:

\begin{equation}\label{criterion}
    \Delta \gg \alpha_0\gamma_2\left(\frac{\pi}{L}\right)^2 + \frac{(-b)}{2}(1 + D_{001})|\epsilon_\parallel|,
\end{equation}

\noindent where $b$ is a valence band deformation potential.

In addition to the perturbation scheme, $H$ is also numerically diagonalized by projecting it into the position basis via the substitution $k_z\to-i\partial/\partial z$, in which the $z$-derivative is implemented by finite differences over the simulation domain. A constant mesh grid size of $0.01\,\text{nm}$ is used for every diagonalization. The Matlab \texttt{eigs()} routine is used to retrieve the desired subset of eigenvalues. The Ge Luttinger parameters $\gamma_{1,2,3}$ and deformation potentials are taken from Ref.~\onlinecite{Paul2016}, while the parameter $\kappa$ is taken from Ref.~\onlinecite{Lawaetz1971}. Explicit matrix representations of $H_0$ and $H'$ are presented in Appendix \ref{app:matrices}. See Appendix \ref{app:H0} for additional details on the eigenvalues and eigenvectors of $H_0$.

Let us now test the accuracy of the perturbative model compared to the dispersion given by solving numerically $H$. We take Ge as the QW material with width $L$ and strain $\epsilon_\parallel$ as free parameters. Since Ge has a rather high spin-orbit energy $\Delta = 260\,\text{meV}$,\cite{Polak2017} it is worthwhile to look also at the behavior of the model with approximations \eqref{B*_approx} and \eqref{F_approx}. We also focus on relaxed or compressively strained wells, which always result in a HH-like valence band edge. The calculated fan diagram of the ground HH subband is displayed in Fig.~2a for a $16\,\text{nm}$-thick well with $\epsilon_\parallel = -0.6\%$, similar to the system analyzed in Fig.~1. Assuming finite $\Delta$, the model reproduces perfectly well the numerical fan diagram up to $\sim 2\,\text{T}$, which implies that $6(\kappa - F)$ is a very accurate approximation for the HH g-factor at low fields. As the magnetic field increases, quadratic terms in $B$ become more important and the dispersions eventually cross. The dispersion of a state with spin-up projection in a given spin-split Landau pair always has a bigger curvature than the spin-down one, which can be straightforwardly inferred from the coefficients $n(n + 1)$ and $(n-2)(n-1)$ in \eqref{up} and \eqref{down}. For that reason, a Zeeman crossing cannot occur, at least to second order, if the spin-up state lies closer to the band gap than the spin-down one. Crossing fields are indicated in Fig.~2a for filling factors $\nu = 3$ and $\nu = 5$. The numerical solution of $H$ gives a crossing field $B_c = 7.27\,\text{T}$ for $\nu = 3$, whereas the second order formula (Eq. \eqref{crossing}) gives $B_c^{(2)} = 5.04\,\text{T}$. Here the second order approximation underestimates $B_c$ as it diverges from the numerical dispersion before the crossing. When assuming $\Delta\to\infty$, however, the dispersion diverges less dramatically than its finite SOI counterpart and instead overestimates the crossing field. Assuming an infinite SOI for this particular system turns out to be a good approximation, because the right-hand side of \eqref{criterion} equals $21.2\,\text{meV}$, which is much smaller than spin-orbit gap in Ge.

\begin{figure}[t]
    \centering
    \includegraphics[scale=0.8]{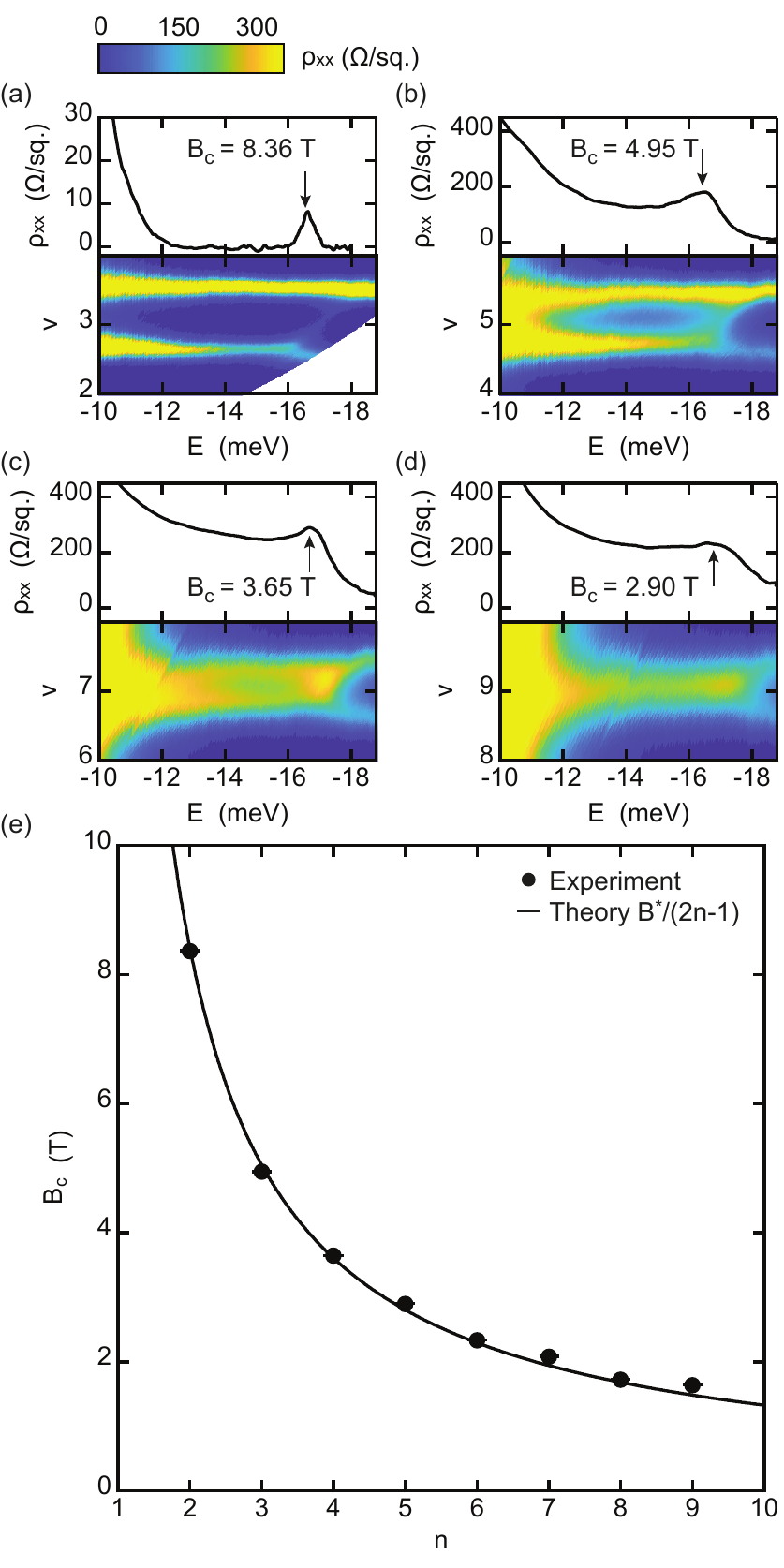}
    \caption{{\bf Experiment vs. Theory.} (a)-(d) $\rho_{xx}$ as a function of filing factor $\nu$ and energy $E$ around the crossings of Zeeman split states. The upper parts of each panel shows a cross-section at odd filling factors $\nu = 3,5,7,9$. (e) Experimental crossing fields (dots) for $\nu = 3, 5, 7, 9, 11, 13, 15, 17$ fitted using Eq. \eqref{crossing} (solid line). The fitting parameter $B^*=25.258\,\text{T}$.}
    \label{fig:exp-vs-theory}
\end{figure}

Fig.~2b depicts the behavior of the crossing field as a function of the well thickness and strain, with and without the assumption of an infinite SOI. The crossing field $B_c$ is well approximated by $B_c^{(2)}$ for a well thickness $> 10\,\text{nm}$ with reduced strain levels, as in our experiments. For narrower and highly strained wells, third or higher perturbative terms become more important. These could be included in the model, but at the cost of extremely cumbersome equations, even with infinite SOI. On the other hand, for $\Delta\to\infty$, $B_c^{(2)}$ misses completely the increase of the crossing field for thin wells, which highlights the explicit role of the SOI strength. This is consistent with criterion \eqref{criterion}~: thin wells increase the right-hand side in \eqref{criterion} as $1/L^2$, thus requiring $\Delta$ to be even larger for this criterion to be satisfied.

\section{Discussion}

From the present model, we see that Zeeman crossings still occur under the assumption of an infinite QW (no barrier effects), an infinite band gap (6-band $k\cdot p$), and even an infinite spin-orbit gap (4-band $k\cdot p$ for HH and LH). Consequently, LH-HH mixing plays a crucial role in the crossing of spin-split states. Our assumptions also imply that structure inversion asymmetry (SIA) has no role in the observed crossing in ZS energy. SIA is indeed suppressed in infinite wells without external electric fields. Thus, Rashba SOI does not have a dominant effect on the value of $B_c$. The role of SOI and strain is, however, more evident in Eqs. \eqref{crossing} and \eqref{B*}. SOI and strain affect $B_c^{(2)}$ mostly through the energy splitting $E^{\text{HH}} - E^\eta$ and the parameter $F$. Compressive strain typically increases $E^{\text{HH}} - E^\eta$, which explains the increase of $B_c$ at higher compressive strain. SOI also increases $E^{\text{HH}} - E^\eta$, mainly through the spin-orbit energy $\Delta$ for $\eta = +$ or through the out-of-plane effective mass for $\eta = -$. At $\Delta = 0$ and any strain, the HH subbands share the same spectrum as the $\eta = +$ or $\eta = -$ states. Eq. \eqref{B*} then gives $B^* = 0$ hence no Zeeman crossing occurs. SOI lifts this degeneracy between HH and $\eta$ states and thus allows the existence of Zeeman crossings.

The experimental observation of Zeeman crossings are further highlighted by plotting portions of the fan diagram from Fig.~1b as a function of energy and filling factor (Fig.~3a-d). The upper part of each panel shows the $\rho_{xx}$ as a function of the energy $E$ at odd-integer values of filling factors from $\nu = 3$ to $9$. Fingerprints of Zeeman crossing are observed for filing factors up to $\nu = 17$. In addition to describing the crossings in Zeeman split states, the theoretical framework described above also allows a straightforward evaluation of several parameters. First, we fit the crossing fields extracted from Fig.~3a-d ($\nu = 3,5,...,17$) with Eq. \eqref{crossing} using $B^*$ as the sole fitting parameter. This yields $B^* = 25.258\,\text{T}$ and the crossing fields obtained from Eq. \eqref{crossing} match the experimental values with a relative error $<4\%$ for $\nu = 3,5,7,9,11$ and $<10\%$ for $\nu = 13,15,17$ (Fig.~3e). Zeeman crossings also approach a fixed energy value as $\nu$ increases, as demonstrated in Eq. \eqref{deltaE}. From Fig.~1(b), we have $\Delta E \approx 17\,\text{meV}$. Knowing $B^*$ and $\Delta E$ gives the value of $F$, leading to HH effective mass and weak-field g-factor. A rearrangement of Eq. \eqref{deltaE} gives

\begin{equation}\label{F_value}
  F = \frac{4}{9}\left(\gamma_1 + \gamma_2 - \frac{\Delta E}{\mu_{\text{B}}B^*}\right) - \frac{\kappa}{3} \approx 1.52.
\end{equation}

From Eqs. \eqref{g-factor} and \eqref{F_value}, we extract $g^* = 11.35$, which is close to $12.9$ obtained by solving $H$ numerically. An expression for the subband-edge HH in-plane effective mass $m^*$ involving the parameter $F$ can also be derived by inserting Eq. (5) from Ref.~\onlinecite{Drichko2018} into Eq. \eqref{g-factor}~: $m^* / m_0 = \left(\gamma_1 + \gamma_2 - 3F\right)^{-1} \approx 0.077$. This value is also close to those reported in the literature at similar hole density.\cite{Lodari2019,Terrazos2018} A close relation exists between the crossing fields, the HH g-factor and the HH-$\eta$ splitting (Eqs. \eqref{B*} and \eqref{B*_approx}). Knowing two of these quantities is enough to obtain the third. For the system described in Fig.1, the criterion \eqref{criterion} is also satisfied, thus the HH-LH splitting is found directly from Eq. \eqref{B*_approx}~:

\begin{equation}\label{HH-LH}
  E^{\text{HH}} - E^{\text{LH}} = \frac{6\left(\gamma_2 + \gamma_3\right)^2\mu_{\text{B}}B^*}{g^*} \approx 76.0\,\text{meV}.
\end{equation}

A numerical solution of $H$ yields a HH-LH splitting of $62.8\,\text{meV}$. This value does not change significantly when an effective out-of-plane electric field is introduced in $H$. This is expected from square QWs whose HH-LH splitting is dominated by strain and quantum confinement.\cite{Moriya2014} For that reason, we assume that the HH-LH splitting does not change with hole concentration, or applied gate voltages. From the HH-LH splitting energy (Eq. \eqref{HH-LH}), one can finally estimate the cubic Rashba coefficient $\alpha_3$~:

\begin{equation}
  \alpha_3 = \frac{e\alpha_0^2\gamma_3}{12\left(\gamma_2 + \gamma_3\right)^3}\left(\frac{g^*}{\mu_{\text{B}}B^*}\right)^2 \approx 4.25\times10^5\,e\,\text{\AA}^4,
\end{equation}

where $e$ is the elementary charge. $\alpha_3$ appears in the cubic Rashba SOI Hamiltonian of HH states\cite{Moriya2014}~: $H_3 = \beta_3i(k_-^3\sigma_+ - k_+^3\sigma_-)$, where $k_\pm = k_x \pm ik_y$ and $\sigma_\pm = (\sigma_x \pm i\sigma_y)/2$ with $\sigma_{x,y}$ the Pauli spin matrices, and $\beta_3 = \alpha_3 E_z$, with $E_z = ep/\epsilon$ the effective out-of-plane electric field in the accumulation mode 2DHG,\cite{Winkler2003} $p$ the hole density and $\epsilon$ the Ge dielectric constant. The obtained $\alpha_3$ is almost twice as large as the one obtained for Ge QW in Ref.~\onlinecite{Moriya2014}, which had a bigger HH-LH splitting of $110\,\text{meV}$. As mentioned above, we expect $\alpha_3$ to be independent of the gate voltage or hole concentration, since it depends mostly on the HH-LH splitting. The Zeeman crossings appear at a density $p\sim 6.1\times 10^{11}\,\text{cm}^{-2}$, corresponding to $E_z \approx 6.8\times 10^{-4}\,\text{V}\,\text{\AA}^{-1}$ (by taking $\epsilon = 16.2\epsilon_0$ for Ge), which yields $\beta_3 \approx 290\,\text{eV}\,\text{\AA}^3$. Note that $\alpha_3$ or $\beta_3$ are hitherto hard to measure in these high mobility systems with established methodologies~: weak anti-localization measurements are impractical due to the small characteristic transport field $B_L$ associated with $\mu$m-scale mean free paths\cite{Hikami1980,Iordanskii1994}~; Shubnikov-de Haas oscillations lack sufficient spectral resolution before onset of ZS to resolve the beatings associated with spin-split subbands.\cite{Hendrickx2018}

\section{Conclusion}
In summary, Zeeman energy crossing of HH states is observed in a Ge 2DHG under out-of-plane magnetic fields and discussed within a perturbative model describing the hole dispersion. Only second order perturbation in the magnetic field is necessary to describe the crossing in which SOI emerges as an essential feature. However, our analysis indicates that SIA has no effective role. Additionally, this analysis also provides a straightforward framework to evaluate several physical parameters defining the hole states from simple magnetotransport measurements. Crucially, the detailed knowledge of parameters such as the effective g-factor, the in-plane effective mass, and the cubic Rashba coefficient of the underlying material platform will provide the necessary input to further advance design and modelling of hole spin qubits and other hole-based quantum devices.

\subsection*{Acknowledgment}

O.~M. acknowledges support from NSERC Canada (Discovery, SPG, and CRD Grants), Canada Research Chairs, Canada Foundation for Innovation, Mitacs, PRIMA Qu\'ebec, Defence Canada (Innovation for Defence Excellence and Security, IDEaS), and NRC Canada (New Beginnings Initiative). G.~S. and M.~L. acknowledge financial support from The Netherlands Organization for Scientific Research (NWO).

\subsection*{Data availability}

Datasets supporting the findings of this study are available at 10.4121/uuid:c64b0509-2247-4d51-adc0-90e361b928a4

\appendix
\section{Hamiltonian matrix representation}\label{app:matrices}

The matrix representation of $H$ is presented in the following $\Ket{j,m}$ angular momentum basis\cite{Winkler2003}~:

\begin{equation*}
\left\{\Ket{\frac{3}{2},\frac{3}{2}},\Ket{\frac{3}{2},\frac{1}{2}},\Ket{\frac{3}{2},-\frac{1}{2}},\Ket{\frac{3}{2},-\frac{3}{2}},\Ket{\frac{1}{2},\frac{1}{2}},\Ket{\frac{1}{2},-\frac{1}{2}}\right\}
\end{equation*}

The magnetic field-free Hamiltonian $H_0$ is

\begin{widetext}

\begin{equation*}
  \begin{split}
    H_0 &= -\alpha_0\begin{bmatrix}
    \left(\gamma_1-2\gamma_2\right)k_z^2 & 0 & 0 & 0 & 0 & 0 \\
    & \left(\gamma_1+2\gamma_2\right)k_z^2 & 0 & 0 & -\sqrt{8}\gamma_2k_z^2 & 0 \\
    &  & \left(\gamma_1+2\gamma_2\right)k_z^2 & 0 & 0 & \sqrt{8}\gamma_2k_z^2 \\
    & \dag &  & \left(\gamma_1-2\gamma_2\right)k_z^2 & 0 & 0 \\
    &  &  &  & \gamma_1k_z^2 + \Delta/\alpha_0 & 0 \\
    &  &  &  &  & \gamma_1k_z^2 + \Delta/\alpha_0 \\
    \end{bmatrix} \\
    &+ \mathbb{1}_{6\times 6}(2 - D_{001})a_v\epsilon_\parallel + \begin{bmatrix}
    1 & 0 & 0 & 0 & 0 & 0 \\
    & -1 & 0 & 0 & \sqrt{2} & 0 \\
    &  & -1 & 0 & 0 & -\sqrt{2} \\
    & \dag &  & 1 & 0 & 0 \\
    &  &  &  & 0 & 0 \\
    &  &  &  &  & 0 \\
    \end{bmatrix}(1 + D_{001})b\epsilon_\parallel,
  \end{split}
\end{equation*}

\end{widetext}

where $a_v$ and $b$ are deformation potentials. Its eigenstates and eigenvalues are described in Appendix \ref{app:H0}. For perpendicular-to-plane magnetic fields it is convenient to write $k_x$ and $k_y$ in terms of the ladder operator $a$~:

\begin{align*}
  k_x &= \frac{1}{\sqrt{2}\lambda}\left(a + a^\dag\right) & k_y &= \frac{i}{\sqrt{2}\lambda}\left(a - a^\dag\right),
\end{align*}

where the magnetic length $\lambda = \sqrt{\hbar/eB}$, $e$ being the elementary charge. Also, $[a,a^\dag] = 1$, $a\Ket{N} = \sqrt{N}\Ket{N-1}$ and $a^\dag a\Ket{N} = N\Ket{N}$, where $N$ is an integer. In the axial approximation the vector

\begin{equation*}
  \begin{bmatrix*}[r]
    \Ket{N-1}\Ket{l}_{3/2,3/2} \\
    \Ket{N}\Ket{l}_{3/2,1/2} \\
    \Ket{N+1}\Ket{l}_{3/2,-1/2} \\
    \Ket{N+2}\Ket{l}_{3/2,-3/2} \\
    \Ket{N}\Ket{l}_{1/2,1/2} \\
    \Ket{N+1}\Ket{l}_{1/2,-1/2} \\
  \end{bmatrix*}
\end{equation*}

is an eigenstate of $H$, where $\Braket{z|l}_{j,m}$ is the spatial envelope function of the hole component with angular momentum $\Ket{j,m}$ and subband index $l$. This ansatz allows to write $H$ as a function of the quantum numbers $N$ and to eliminate the ladder operators $a$.\cite{Luttinger1956} The perturbation $H'$ takes the form (with the electron g-factor $g_0 = 2$)~:

\begin{widetext}

\begin{align*}
  H' = -\frac{\alpha_0}{\lambda^2}&\left[\begin{matrix}
    (2N-1)\gamma_+ + 3\kappa & -2\lambda\sqrt{6N}\gamma_3k_z & -\sqrt{3N(N+1)}\tilde{\gamma} \\
    & (2N+1)\gamma_- + \kappa & 0 \\
    & & (2N+3)\gamma_- - \kappa \\
    & \dag & \\
    & & \\
    & & \\
    \end{matrix}\right. \dots \\
    &\left.\begin{matrix}
    0 & 2\lambda\sqrt{3N}\gamma_3k_z & \sqrt{6N(N+1)}\tilde{\gamma} \\
    -\sqrt{3(N+1)(N+2)}\tilde{\gamma} & \sqrt{2}[(2N+1)\gamma_2+\kappa+1] & -6\lambda\sqrt{N+1}\gamma_3k_z \\
    2\lambda\sqrt{6(N+2)}\gamma_3k_z & -6\lambda\sqrt{N+1}\gamma_3k_z & -\sqrt{2}[(2N+3)\gamma_2-\kappa-1] \\
    (2N+5)\gamma_+ -3\kappa & -\sqrt{6(N+1)(N+2)}\tilde{\gamma} & 2\lambda\sqrt{3(N+2)}\gamma_3k_z \\
    & (2N+1)\gamma_1 + 2\kappa + 1 & 0 \\
    & & (2N+3)\gamma_1 - 2\kappa - 1 \\
    \end{matrix}\right]
\end{align*}

\end{widetext}

where $N\geq -2$, $\gamma_\pm = \gamma_1 \pm \gamma_2$ and $\tilde{\gamma} = \gamma_2 + \gamma_3$. Note that $\alpha_0/\lambda^2 = \mu_{\text{B}}B$.

\section{Eigenvalues and eigenvectors of \texorpdfstring{$H_0$}{H0}}\label{app:H0}

If $B = 0$ (and $k_x = k_y = 0$) the subbands are either pure HH states or pure spin-1/2 states (LH-SO superposition). The subband eigenstates are

\begin{align}
  \Ket{\text{HH},\sigma,l} &= \Ket{\frac{3}{2},\frac{3\sigma}{2}}\Ket{l} \\
  \Ket{\eta,\sigma,l} &= \left(l_l^\eta\Ket{\frac{3}{2},\frac{\sigma}{2}} + \sigma s_l^\eta\Ket{\frac{1}{2},\frac{\sigma}{2}}\right)\Ket{l},
\end{align}

where $\sigma = \{+,-\}$ is the pseudo-spin index (spin-up -down respectively). We have

\begin{align}
  l_l^\eta &= \frac{\xi_- + \delta_{-\eta}}{\sqrt{\left(\xi_- + \delta_{-\eta}\right)^2 + 8\xi_-^2}}, &s_l^\eta &= \frac{-\sqrt{8}\xi_-}{\sqrt{\left(\xi_- + \delta_{-\eta}\right)^2 + 8\xi_-^2}},
\end{align}

with

\begin{gather}
  \delta_\pm = \frac{\Delta}{2} \pm \sqrt{\left(\xi_- + \Delta/2\right)^2 + 8\xi_-^2}, \\
    \begin{split}
    \xi_\pm &= -\alpha_0\left[\left(\frac{1 \pm 1}{2}\right)\gamma_1 + \gamma_2\right]\left(\frac{l\pi}{L}\right)^2 \\
    &+ \left[\left(\frac{1\pm 1}{2}\right)(2 - D_{001})a_v - (1+D_{001})b/2\right]\epsilon_\parallel.
    \end{split}
\end{gather}

The spatial part is

\begin{equation}
  \Braket{z|l} = \sqrt{\frac{2}{L}}\sin\left(\frac{l\pi z}{L}\right).
\end{equation}

The energy spectrum for the HH and $\eta$-states is

\begin{align}
  \begin{split}
  E_l^{\text{HH}} &= -\alpha_0\left(\gamma_1 - 2\gamma_2\right)\left(\frac{l\pi}{L}\right)^2 \\
  &+ \left[(2 - D_{001})a_v + (1+D_{001})b\right]\epsilon_\parallel
  \end{split} \\
  E_l^\eta &= \xi_+ - \delta_\eta.
\end{align}

Infinite SOI regime is reached when $\Delta \gg |\xi_-|$. Under compressive strain this expands to

\begin{equation}\label{app:criterion}
  \Delta \gg \alpha_0\gamma_2\left(\frac{l\pi}{L}\right)^2 + \frac{(-b)}{2}(1 + D_{001})|\epsilon_\parallel|.
\end{equation}

Assuming $\Delta\to\infty$ is a good approximation only if $\Delta$ satisfies criterion \eqref{app:criterion}. The square root in $\delta_\pm$ can then be eliminated by a Taylor expansion, and the following results immediately follow~:

\begin{align*}
  \delta_- &= -\xi_-, & \delta_+ &= \Delta + \xi_-, \\
  l_l^- &= 1, & s_l^- &= 0, \\
  l_l^+ &= 0, & s_l^+ &= 1.
\end{align*}

Consequently,

\begin{gather}
  \Ket{-,\sigma,l} \rightarrow \Ket{\text{LH},\sigma,l} = \Ket{\frac{3}{2},\frac{\sigma}{2}}\Ket{l} \\
  \Ket{+,\sigma,l} \rightarrow \Ket{\text{SO},\sigma,l} = \Ket{\frac{1}{2},\frac{\sigma}{2}}\Ket{l} \\
  \begin{split}
  E_l^- \rightarrow E_l^{\text{LH}} &= -\alpha_0\left(\gamma_1 + 2\gamma_2\right)\left(\frac{l\pi}{L}\right)^2 \\
  &+ \left[(2 - D_{001})a_v - (1+D_{001})b\right]\epsilon_\parallel
  \end{split} \\
  E_l^+ \rightarrow E_l^{\text{SO}} = -\alpha_0\gamma_1\left(\frac{l\pi}{L}\right)^2 - \Delta + (2 - D_{001})a_v\epsilon_\parallel,
\end{gather}

corresponding to a pure LH and pure SO spectrum.

\end{document}